\documentclass{PoS}

\newcommand{\HI}{H\,{\sc i}{}}

\title{An Overview of the MHONGOOSE Survey: Observing Nearby Galaxies with MeerKAT}

\ShortTitle{An Overview of the MHONGOOSE Survey}

\author{\speaker{W.J.G. de Blok}$^{1,2,3}$\\
  E-mail: \email{blok@astron.nl}}

\author{E.A.K. Adams$^{1,3}$,
P. Amram$^4$,
E. Athanassoula$^4$,
I. Bagetakos$^{1,3}$,
C. Balkowski$^5$,
M.A. Bershady$^6$ ,
R. Beswick$^7$,
F. Bigiel$^8$ ,
S.-L. Blyth$^2$,
A. Bosma$^4$,
R.S. Booth$^9$,
A. Bouchard$^{10}$,
E. Brinks$^{11}$,
C. Carignan$^2$,
L. Chemin$^{12}$,
F. Combes$^5$,
J. Conway$^{13}$,
E.C. Elson$^{14}$,
J. English$^{15}$,
B. Epinat$^4$,
B.S. Frank$^2$,
J. Fiege$^{15}$,
F. Fraternali$^{16,3}$ ,
J.S. Gallagher$^{6}$,
B.K. Gibson$^{17}$,
G. Heald$^{18}$,
P.A. Henning$^{19}$ ,
B.W. Holwerda$^{20}$,
T.H. Jarrett$^2$,
H. Jerjen$^{21}$,
G.I. J\'ozsa$^{22,23,24}$,
M. Kapala$^2$,
H.-R. Kl\"ockner$^{25}$ ,
B.S. Koribalski$^{18}$,
R.C. Kraan-Korteweg$^2$,
S. Leon$^{26}$,
A. Leroy$^{27}$,
S.I. Loubser$^{28}$ ,
D.M. Lucero$^3$,
S.S. McGaugh$^{29}$,
G.R. Meurer$^{30}$,
M. Meyer$^{30}$,
M. Mogotsi$^{31}$,
B. Namumba$^2$ ,
S-H. Oh$^{32}$,
T.A. Oosterloo$^{1,3}$,
D.J. Pisano$^{33,34}$ ,
A. Popping$^{30}$,
S. Ratcliffe$^{22}$,
J.A. Sellwood$^{35}$,
E. Schinnerer$^{36}$,
A.C. Schr\"oder$^{31}$,
K. Sheth$^{37}$,
M.W.L. Smith$^{38}$,
A. Sorgho$^{2}$,
K. Spekkens$^{39}$,
S. Stanimirovic$^6$,
K. van der Heyden$^2$,
W. van Driel$^5$,
L. Verdes-Montenegro$^{40}$,
F. Walter$^{36}$,
T. Westmeier$^{30}$,
E. Wilcots$^6$,
T. Williams$^{31}$,
O.I. Wong$^{30}$,
P.A. Woudt$^2$,
A. Zijlstra$^{41}$\\
\llap{$^1$} ASTRON Netherlands Institute for Radio Astronomy, the Netherlands\\
\llap{$^2$} Univ. of Cape Town, South Africa\\
\llap{$^3$} Kapteyn Institute, Univ. of Groningen, the Netherlands\\
\llap{$^4$} Lab. Astroph. Marseille France\\
\llap{$^5$} Obs. de Paris, France\\
\llap{$^6$} Univ. of Wisconsin, USA \\
\llap{$^7$} Jodrell Bank, UK\\
\llap{$^8$} Univ. Heidelberg, Germany \\
\llap{$^9$} HartRAO, South Africa\\
\llap{$^{10}$} McGill Univ., Canada\\
\llap{$^{11}$} Univ. of Hertfordshire, UK\\
\llap{$^{12}$} Univ. de Antofagasta, Chile\\
\llap{$^{13}$} Chalmers Univ., Sweden \\
\llap{$^{14}$} Univ. of Western Cape, South Africa\\
\llap{$^{15}$} Univ. Manitoba, Canada\\
\llap{$^{16}$} Univ. Bologna, Italy \\
\llap{$^{17}$} E.A. Milne Center for Astrophysics, Univ. of Hull, UK \\
\llap{$^{18}$} CSIRO Astronomy and Space Science, ATNF, Australia\\
\llap{$^{19}$} Univ. New Mexico, USA\\
\llap{$^{20}$} Univ. Louisville, USA\\
\llap{$^{21}$} RSAA, ANU, Australia \\
\llap{$^{22}$} SKA-SA, South Africa\\
\llap{$^{23}$} Rhodes Univ., South Africa\\
\llap{$^{24}$} Argel\"ander-Institut f\"ur Astronomy, Germany\\
\llap{$^{25}$} MPIfR, Bonn, Germany\\
\llap{$^{26}$} ESO, Chile\\
\llap{$^{27}$} Ohio State Univ., USA \\
\llap{$^{28}$} North-West Univ., South Africa\\
\llap{$^{29}$} Case Western Reserve Univ., USA\\
\llap{$^{30}$} ICRAR/UWA, Australia\\
\llap{$^{31}$} SAAO, South Africa\\
\llap{$^{32}$} KASI, South Korea\\
\llap{$^{33}$} West Virginia Univ., USA\\
\llap{$^{34}$} Center for Gravitational Wave and Cosmology, Morgantown, WV, USA\\
\llap{$^{35}$} Steward Observatory, USA\\
\llap{$^{36}$} MPIA, Heidelberg, Germany \\
\llap{$^{37}$} NASA, Washington DC, USA\\
\llap{$^{38}$} Cardiff Univ., UK\\
\llap{$^{39}$} RMC, Canada\\
\llap{$^{40}$} IAA, Spain\\
\llap{$^{41}$} Univ. Manchester, UK
}

\abstract{MHONGOOSE is a deep survey of the neutral hydrogen
  distribution in a representative sample of 30 nearby disk and dwarf
  galaxies with \HI\ masses from $\sim 10^6$ to $\sim 10^{11}\,
  M_{\odot}$, and luminosities from $M_R \sim −12$ to $M_R \sim
  -22$. The sample is selected to uniformly cover the available range
  in $\log(M_{HI})$. Our extremely deep observations, down to
  \HI\ column density limits of well below $10^{18}$ cm$^{-2}$ --- or
  a few hundred times fainter than the typical \HI\ disks in galaxies
  --- will directly detect the effects of cold accretion from the
  intergalactic medium and the links with the cosmic web.  These
  observations will be the first ever to probe the very low-column
  density neutral gas in galaxies at these high
  resolutions.\\ Combination with data at other wavelengths, most of
  it already available, will enable accurate modeling of the
  properties and evolution of the mass components in these galaxies
  and link these with the effects of environment, dark matter
  distribution, and other fundamental properties such as halo mass and
  angular momentum.\\ MHONGOOSE can already start addressing some of
  the SKA-1 science goals and will provide a comprehensive inventory
  of the processes driving the transformation and evolution of
  galaxies in the nearby universe at high resolution and over 5 orders of
  magnitude in column density.  It will be a Nearby Galaxies Legacy
  Survey that will be unsurpassed until the advent of the SKA, and can
  serve as a highly visible, lasting statement of MeerKAT's
  capabilities.}

\FullConference{MeerKAT Science: On the Pathway to the SKA\\
		 25-27 May, 2016\\
		 Stellenbosch, South Africa}

\begin{document}
\section{Introduction}

One of the Key Science Questions for the Square Kilometre Array (SKA)
is ``How do galaxies assemble and evolve?''  The SKA will be able to
trace the gradual transformation from primordial neutral hydrogen (\HI)
into galaxies over cosmic time. However, direct, detailed and
  resolved observations of the sub-kpc-scale physical processes that
cause this transformation, taking place both inside and around these
evolving galaxies, will probably stay beyond our reach --- even with
the SKA --- for a large span of cosmic time due to resolution and
sensitivity limitations.

The only place where a comprehensive survey of the ``Galactic
ecosystem'' can be made is the nearby universe; only locally can we
study, in detail, the ``baryon cycle'', i.e., the flow of gas into
galaxies, its physical conditions, its transformation into stars, and
how it, in turn, is affected by feedback.  Resolved \HI\ observations
will tell us how galaxies acquire their gas, how star formation
is sustained and, ultimately, how the dark and visible matter together
determine and regulate the evolution of galaxies.

Local galaxies are the ``fossil records'' of the distant,
high-redshift galaxies, and provide a wealth of information that will
help refine models of galaxy formation and evolution.  They provide
the foundations on which studies of higher redshift galaxies must be
built.

In 2010, time was allocated on MeerKAT to carry out
\emph{MHONGOOSE}\footnote{MeerKAT \HI\ Observations of Nearby Galactic
  Objects; Observing Southern Emitters}, a deep \HI\ survey of 30 nearby
galaxies.  The MHONGOOSE observations aim to reach a $3\sigma$ column
density limit of $7.5 \cdot 10^{18}$ cm$^{-2}$, at a resolution of
$30''$ and integrated over 16 km s$^{-1}$ (roughly the width of the \HI\ 
line). At 90$''$ resolution the corresponding $3\sigma$ limit is $5.5
\cdot 10^{17}$ cm$^{-2}$.

The large number of
short baselines of MeerKAT will efficiently detect low column
density material. Compared to telescopes like the VLA or WSRT, MeerKAT
can in a single pointing map the \HI\ twice as far out into a galaxy's
halo, providing information on evolutionary processes away from the
star forming disks.

MeerKAT will also be the most efficient telescope for producing
detailed maps of the high-resolution ($\sim 6''$) \HI\ distribution and
kinematics within the disks of nearby galaxies, combining the baseline
distributions of multiple existing (B, C, D) and hypothetical (E) JVLA
configurations in one single array.

Previous surveys of nearby galaxies, such as THINGS \cite{walter08}
and HALOGAS \cite{heald11}, have concentrated on either obtaining a
high spatial resolution or a high column density sensitivity ---
neither THINGS nor HALOGAS achieve both. Thanks to MeerKAT's
combination of exquisite column density sensitivity, high spatial
resolution and large field of view, MHONGOOSE will be the first survey
that does not suffer from these limitations and that will therefore
provide information on the processes driving the transformation and
evolution of galaxies in the nearby universe at high resolution and to
low column densities.

Specifically, it will be possible to investigate the low
column-density \HI, from the outskirts of the star-forming disks out
into the far reaches of the dark matter halo. The observations,
sensitive to column densities some two to three orders of magnitude
lower than found in the main disk, will yield clues on gas flows in
and out of the disk, accretion from the intergalactic medium (IGM),
the fuelling of star formation, the connection with the cosmic web and
even the possible existence of low-mass cold dark matter (CDM)
halos. The higher resolution made possible by MeerKAT will resolve
many of these phenomena, thus enabling a more detailed study of their
internal structure, something not possible with previous surveys.

In the 2009/2010 proposal round, one of the projects submitted focused
on the properties and evolution of magnetic fields in (amongst others)
nearby galaxies (MeerQUITTENS; Bolton et al). While that project was
not allocated survey time on MeerKAT, the team was encouraged to make
use of commensal observations to incorporate MeerQUITTENS questions
into the MeerKAT science programme. The scientific goals of
MeerQUITTENS to determine the detailed 3D structure of gas and
magnetic fields on sub-kpc scales as well as the relationship between
magnetic fields and star formation is still highly relevant today, and
MeerKAT promises uniquely powerful leverage on those questions. For
these reasons we are incorporating a description of some of the
relevant polarization science in this paper.

\section{Science Questions}

The science topics that will be addressed by MHONGOOSE are:
\begin{itemize}
\item the importance and effects of cold gas accretion;
\item detection of the cosmic web;
\item the relation between gas and star formation;
\item the relation between dark and baryonic matter;
\item the distribution of dark matter within galaxies;
\item structure, strength and dynamical importance of magnetic fields.
\end{itemize}

A summary of the scientific background of some of these topics is
given below.

\subsection{Accretion}

\begin{figure}
\centering
\includegraphics[width=0.5\linewidth]{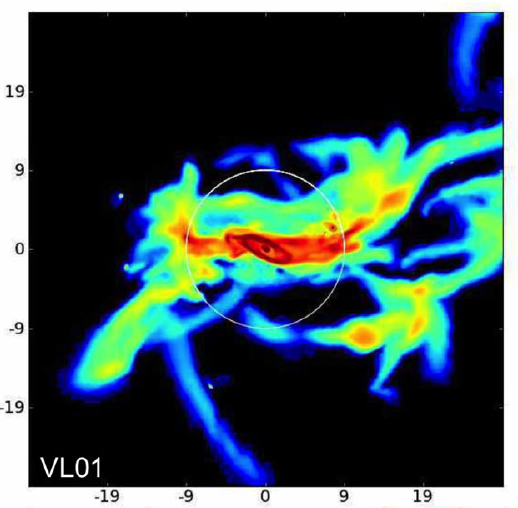}
\caption{Visual impression of the morphology of simulated
  cold accretion features. The small dark brown ring in the center
  represents the main galaxy disk. The figure measures $\sim 60$ kpc
  on the side, the circle has a radius of $\sim 10$ kpc. This area
  would fit within one MeerKAT primary beam for distances $D>4$
  Mpc. Figure taken from \cite{danovich15}.}\label{fig:coldacc}
\end{figure}

In the inner regions of spirals, time scales for consumption of gas by
star formation are much smaller than a Hubble time, even though the
star formation rate has been approximately constant over most of that
time (e.g., \cite{bigiel11}). Galactic disks can, in principle, be
replenished by accreting gas-rich companion galaxies, but the slope of
the \HI\ mass function is not steep enough for small companions to
supply larger galaxies with a substantial amount of gas for a
sufficiently long time. This implies that spirals have to accrete
directly from the IGM.

The presence of cold gas in the halos of our Milky Way and other
galaxies has been known for some time (see, e.g., \cite{wakker97,
  oosterloo07, sancisi08, heald11}). Some of this halo \HI\ is likely to
be part of a star formation driven ``galactic fountain''
\cite{shapiro76}. This is suggested by the observation that some of
the halo \HI\ has a similar projected radial distribution to the star
formation in the disk and that it has disk-like kinematics: rotating
but lagging behind the main disk (see, e.g., \cite{fraternali01}).

However, some of the \HI\ complexes found outside the disks are
counter-rotating with respect to the disk, so cannot have originated
in it.  Numerical simulations (e.g., \cite{keres05}) predict that
``fingers'' of cooler gas from the IGM can penetrate the hot halos
surrounding galaxies and deposit gas into the disk. This process is
called ``cold accretion''.  Figure \ref{fig:coldacc} gives a visual
impression of typical cold accretion features around simulated
galaxies. It is in the context of this cold accretion that the study
of \HI\ halos of galaxies is relevant: it could provide direct observations
of the accretion of gas onto galaxies and forms a strong observational
test for models of galaxy evolution.

\begin{figure}
\centering\includegraphics[width=0.9\columnwidth]{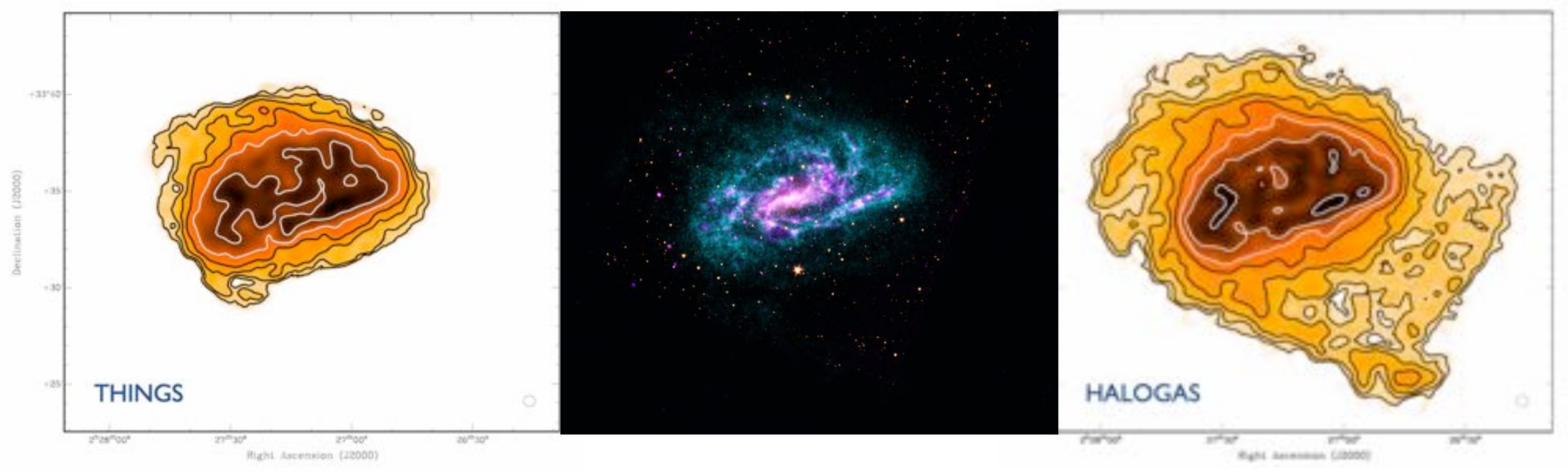} 
\caption{Comparison of shallow THINGS and deep HALOGAS
  observations of NGC 925 at the same spatial resolution. Left: The \HI\ 
  distribution of NGC 925 from THINGS, convolved to the WSRT
  resolution. Lowest contour at $9 \cdot 10^{19}$ cm$^{-2}$, with each
  contour double the previous value.  Center: False-color image of the
  baryonic components. Blue shows the \HI\ distribution derived from
  THINGS, cyan the distribution of young stars derived from GALEX
  data. Right: The \HI\ distribution as observed by the deep WSRT
  HALOGAS survey. Lowest contour $1.8 \cdot 10^{19}$ cm$^{-2}$ with
  each contour double the previous value.\label{fig:halogas}}
\vspace{-0.5cm}
\end{figure}

The current state-of-the art survey of these \HI\ halos is the WSRT
HALOGAS project \cite{heald11}. It has mapped $22$ disk galaxies down
to a column density limit of $\sim 10^{19}$ cm$^{-2}$, i.e., an order
of magnitude lower than typically found in the main \HI\ disks.  The
first results of HALOGAS indicate that some galaxies have extended \HI\
emission at these low levels (see Fig.\ \ref{fig:halogas}), while
others do not: extensive \HI\ halos have been detected in about 12 of
the 22 galaxies observed. It is possible that some of this gas is
related to star formation and galactic fountain processes, but as
discussed above, accretion cannot be excluded. The upper limit on the
cold gas accretion rate as determined by HALOGAS is only $\sim 10\%$
of the current star formation rate in the disk, suggesting most accretion must
occur at lower neutral gas column densities (either because the column
density is truly lower, or because a larger fraction of the gas is
ionised).

The MHONGOOSE observations will probe a factor $\sim 50$ deeper in
column density than HALOGAS and these deep observations will show how
the low column density gas is connected with the cosmic web and where
accretion occurs.  Cold accretion is predicted to be the dominant
process in galaxies with baryonic masses $\log(M_{\rm bar}) < 10.3$
\cite{keres05} corresponding to \HI\ masses below a few times
$10^9\ M_{\odot}$. The latter value is approximately equal to the
average \HI\ mass of the HALOGAS galaxies. To increase the chances of
detecting the direct effects of cold accretion, the MHONGOOSE sample
contains a larger number of galaxies with lower \HI\ masses.

\subsection{Cosmic Web}

The cold accretion process described above delivers gas from the
cosmic web into galaxies. This process is a prediction of
high-resolution numerical models of structure formation (e.g.,
\cite{dave99,crain16}). These predict that most of the baryons at
low redshift are in a warm-hot intergalactic medium (WHIM; T =
10$^5$--10$^7$ K), while 25\% are in the 10$^4$ K diffuse IGM, with
only 25\% condensed in galaxies and their gaseous halos. Due to the
moderately high temperature in the IGM ($>10^4$ K), most of the gas in
the cosmic web is ionised.  To detect the cooler baryons in the cosmic
web, a column density sensitivity of $\sim 10^{17-18}$ cm$^{-2}$ is
required \cite{popping09}. Observationally the presence of cold gas
around galaxies out to radii of at least $\sim 300$ kpc has been
established (e.g., \cite{borisova16}).

MHONGOOSE will have enough sensitivity to reach these column density
values. At a resolution of $90''$, the typical 3$\sigma$ column density
sensitivity of the observations will be $\sim 5 \cdot 10^{17}$
cm$^{-2}$. Stacking the \HI\ profiles will push the effective column
density sensitivity even lower by a factor of several. Pushing radio
technology to the limit is the only way forward: optical telescopes
will for the foreseeable future not be able to directly detect in
emission the ionized gas which the \HI\ traces.

These sensitivities are close to those obtained by very deep
single-dish \HI\ observations. The deepest of these are probably the
observations by \cite{braun04} (using the WSRT as a single dish) of
the low-column density features around and between M31 and M33. In
these observations, the $3\sigma$ limit over 16 km s$^{-1}$ is $1.1
\cdot 10^{17}$ cm$^{-2}$, but with an angular resolution of $\sim
49'$.

A larger collection of very deep observations with the Green Bank
Telescope (GBT) of the HALOGAS and THINGS galaxies has been obtained
by D.J.\ Pisano (in prep.; see \cite{pisano14, deblok14}). These reach
a $3\sigma$ sensitivity of $\sim 6 \cdot 10^{17}$ cm$^{-2}$. However,
sheer column density sensitivity is not enough.  For example, \cite{wolfe13,
  wolfe16} show that the diffuse low column density gas between M31
and M33 observed by \cite{braun04} is resolved in several kpc-sized
clouds when observed at higher spatial resolutions. The $90''$ MeerKAT
beam measures a few kpc at the typical distance of our sample and is
thus very well matched with the expected sizes of the cold accretion
clouds.  It is the powerful combination of column density sensitivity
\emph{and} spatial resolution that makes MeerKAT the ideal instrument
for this work.

\subsection{Gas and star formation}

MHONGOOSE will be able to make several key tests of interstellar
medium (ISM) and star formation physics.  Some of the work on this was
also done by the THINGS survey \cite{walter08} where obtaining a
better understanding of the relation between gas and star formation
was one of the main science goals \cite{leroy08}.  The limited
sensitivity of THINGS constrained these studies, however, to the
optical disk only. The higher sensitivity of the MHONGOOSE
observations, at a similar angular resolution, means these studies can
now be extended to the outer parts of the disks. Molecular gas has
been detected in the far outer parts of disks, so \HI\ certainly changes
phase there, leading to star formation \cite{dessauges14}.

In these outer parts, where \HI\ dominates, the ratio of UV to \HI\ column
density is the key tracer of the timescale and efficiency of star
formation. We will compare this observable to proposed star formation
timescales (e.g., \cite{krumholz07, leroy08, wong09}) and thresholds
(e.g., \cite{schaye04, deblok06, yang07}). With the MeerKAT
observations providing the angular resolution to isolate specific
conditions of the ISM, and the velocity resolution to separate warm
and cold \HI\ \cite{ianja12}, MHONGOOSE promises to be a unique data
set to study star formation in galaxy outskirts.

The MHONGOOSE sample includes edge-on galaxies, ensuring that deep and
sensitive observations will be available for a detailed study of the
vertical distribution of the \HI, associated flaring of the disk, the
presence of gas above the disk as well as galactic fountain-type
outflows due to star formation.

The large range in stellar disk mass in the MHONGOOSE sample will
enable a study of the effect of increasing disk domination on the dark
matter distributions in more massive galaxies. We will infer the
distribution of dark matter and relate this to, e.g., disk mass
density, scale length, disk spin/angular momentum, bulge/disk ratio,
and star formation rate.

\section{Sample Definition and observing time\label{sec:sample}}

\begin{figure}[t]
\centering
\includegraphics[width=\linewidth]{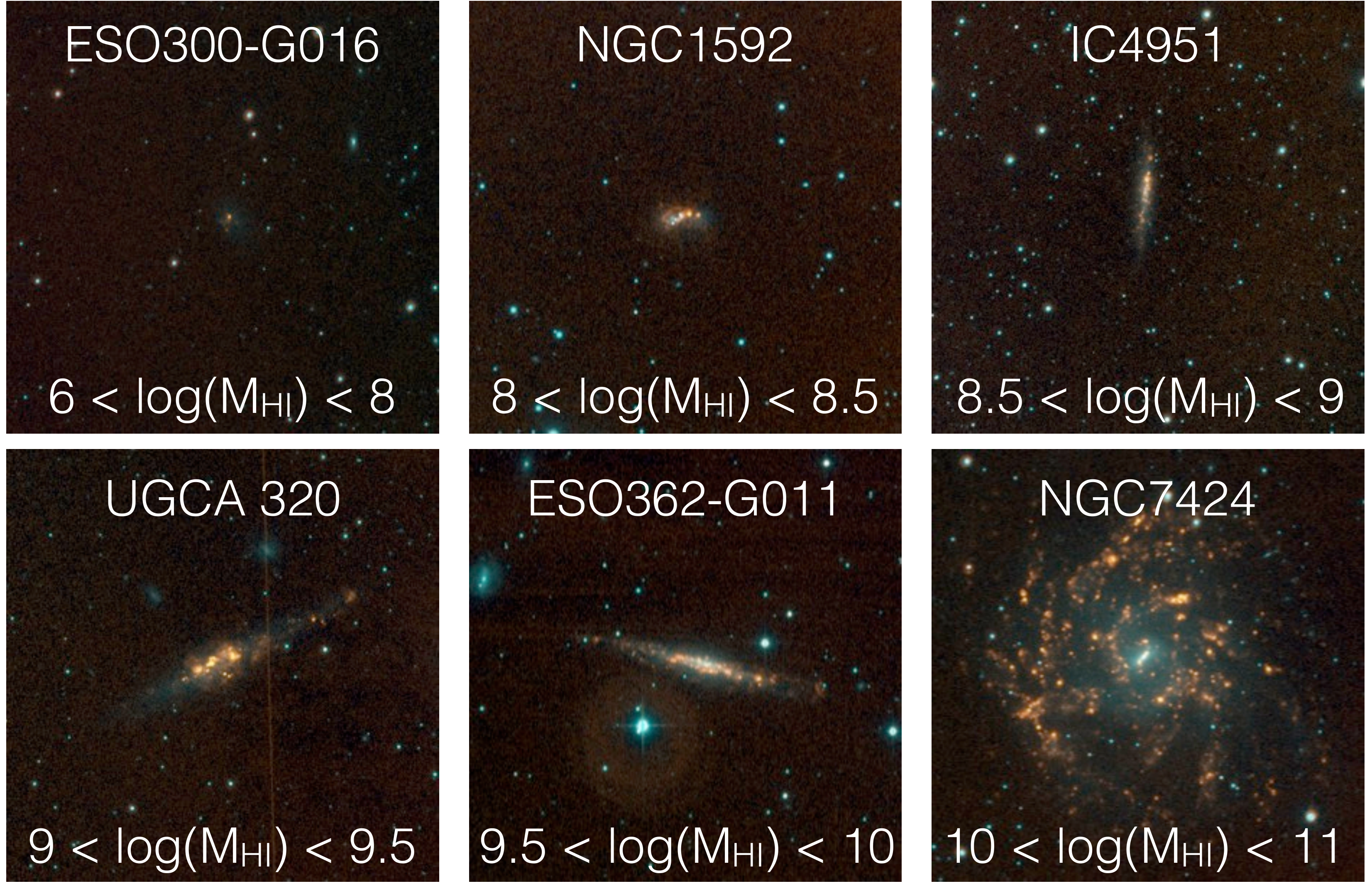}
\caption{SINGG \cite{meurer06} images of one galaxy each from each of
  the 6 \HI\ mass bins. Orange/yellow shows the H$\alpha$ emission, blue
  the optical $R$-band. A full set of images is available on the
  MHONGOOSE website at {\tt http://mhongoose.astron.nl}.
\label{fig:sample}}
\end{figure}

In 2010, time was allocated to study a sample of 30 galaxies, chosen
uniformly from bins in $\log(M_{\rm HI})$ over the range $6 <
\log(M_{HI}) < 11$, thus ensuring a flat distribution in $\log(M_{\rm
  HI})$.  The precursor sample from which our sample has been selected
is based on SINGG (Survey for Ionization in Neutral Gas Galaxies;
\cite{meurer06}).  SINGG targeted $\sim 500$ HIPASS-detected nearby
galaxies, also selected uniformly in bins of log($M_{HI}$).  The SINGG
galaxies were selected to have a HIPASS peak flux $> 50$ mJy, a
galactic latitude $|b| > 30^{\circ}$, a projected distance from the
center of the LMC $> 10^{\circ}$ and a Galactic standard of rest
velocity $> 200$ km s$^{-1}$.  H$\alpha$, optical, infrared and
ultraviolet data are available for the SINGG galaxies.

In selecting the MHONGOOSE galaxies, strongly interacting galaxies and
dense group and cluster environments were avoided, since studies of
isolated galaxies have shown that the gas captured from companion
galaxies and galactic fountain processes (due to star formation and
AGN) are minimized in these environments (AMIGA project: e.g., \cite{espada11a,
  espada11b, lisenfeld07, leon08, sabater12}). The contribution of the
cold gas accretion should thus stand out more prominently this way.
We further limited the sample to $\delta < -10^{\circ}$ and a distance
$D<30$ Mpc (and excluded the MeerKAT Fornax survey region, PI Paolo
Serra).  The distance limit ensures that the beam of MeerKAT at the
highest resolution is always smaller than $\sim 1$ kpc
(comparable to the THINGS resolution).  This left a target list of 88
galaxies. These were divided in 6 bins of $\log(M_{\rm HI})$, and in
each bin 5 galaxies were selected, where each galaxy was selected to
be either edge-on, face-on or with an intermediate inclination of
50--60 degrees.  Face-on allows the best characterization of the
morphology of the ISM, as well as determination of vertical
motions. Edge-on allows an unambiguous characterization of the
vertical structure of the ISM. The intermediate inclination range is
optimal for determining rotation curves and kinematical modeling. Care
was taken that a range in rotation velocity and star formation rate
was covered\footnote{For a more extensive description of
the sample selection see the MHONGOOSE website at
{\tt http://mhongoose.astron.nl}.}.  A selection of SINGG images of the sample galaxies is
shown in Fig.~\ref{fig:sample}.

\begin{figure}
\centering
\includegraphics[width=0.5\linewidth]{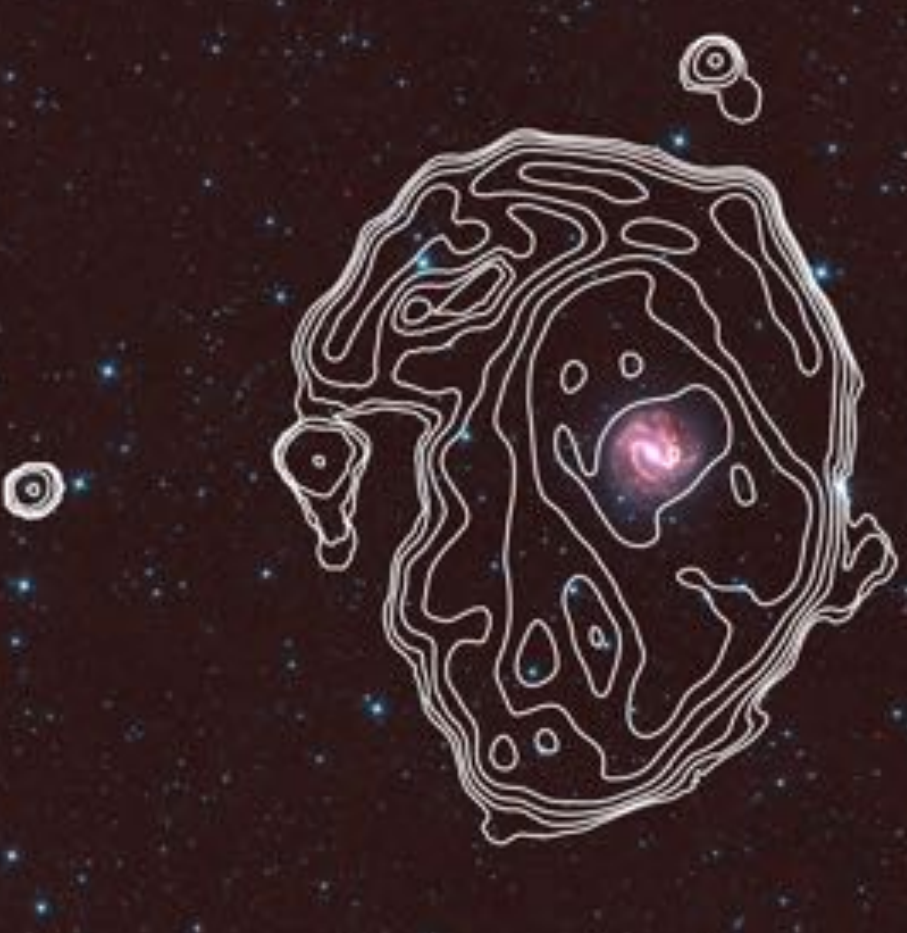}
\caption{KAT-7 3-pointing mosaic of the extended \HI\ distribution of
  M83 \cite{heald16}. Contours show \HI\ column densities and
  start at $5.6 \times 10^{18}$ cm$^{-2}$ increasing by powers of
  1.778. The contours are overlaid on a three-color WISE image. In
  this image, the blue channel is formed from W1 and W2, the green
  channel from W1, W2, and W3, and the red channel from W3 and
  W4. Each channel is displayed using a square-root stretch. The image
  measures 1.5 by 1.5 degrees.}\label{fig:m83}
\vspace{-10pt}
\end{figure}

The desired column density limit for MHONGOOSE is $7.5 \cdot 10^{18}$
cm$^{-2}$ at $3\sigma$ over 16 km s$^{-1}$ at $30''$ resolution.  For
the 2010 MeerKAT parameters, this corresponds to a noise of 0.074 mJy
beam$^{-1}$ per 5 km s$^{-1}$ channel assuming natural
weighting\footnote{The calculation of the column density also includes
  a factor to take into account the increased noise due to tapering;
  cf.\ Fig.\ 4 and Table 3 in the original Call for Large Survey
  Projects \cite{booth09}.}.  With the current, updated MeerKAT
parameters, this noise level is reached after 48$^h$ on-source,
assuming natural weighting. Assuming an overhead of 15\% for set-up
and calibration, results in a total time per galaxy of 55$^h$. The
total observing time to reach this sensitivity for the whole sample
therefore becomes $30 \times 55^h = 1650$ hours.

\section{Comparison with previous surveys}

MHONGOOSE is designed to optimally make use of MeerKAT's unique
capabilities: a high spatial and spectral resolution in combination
with an excellent column density sensitivity and a wide field of view.

\begin{figure}
\centering
\includegraphics[width=0.6\textwidth]{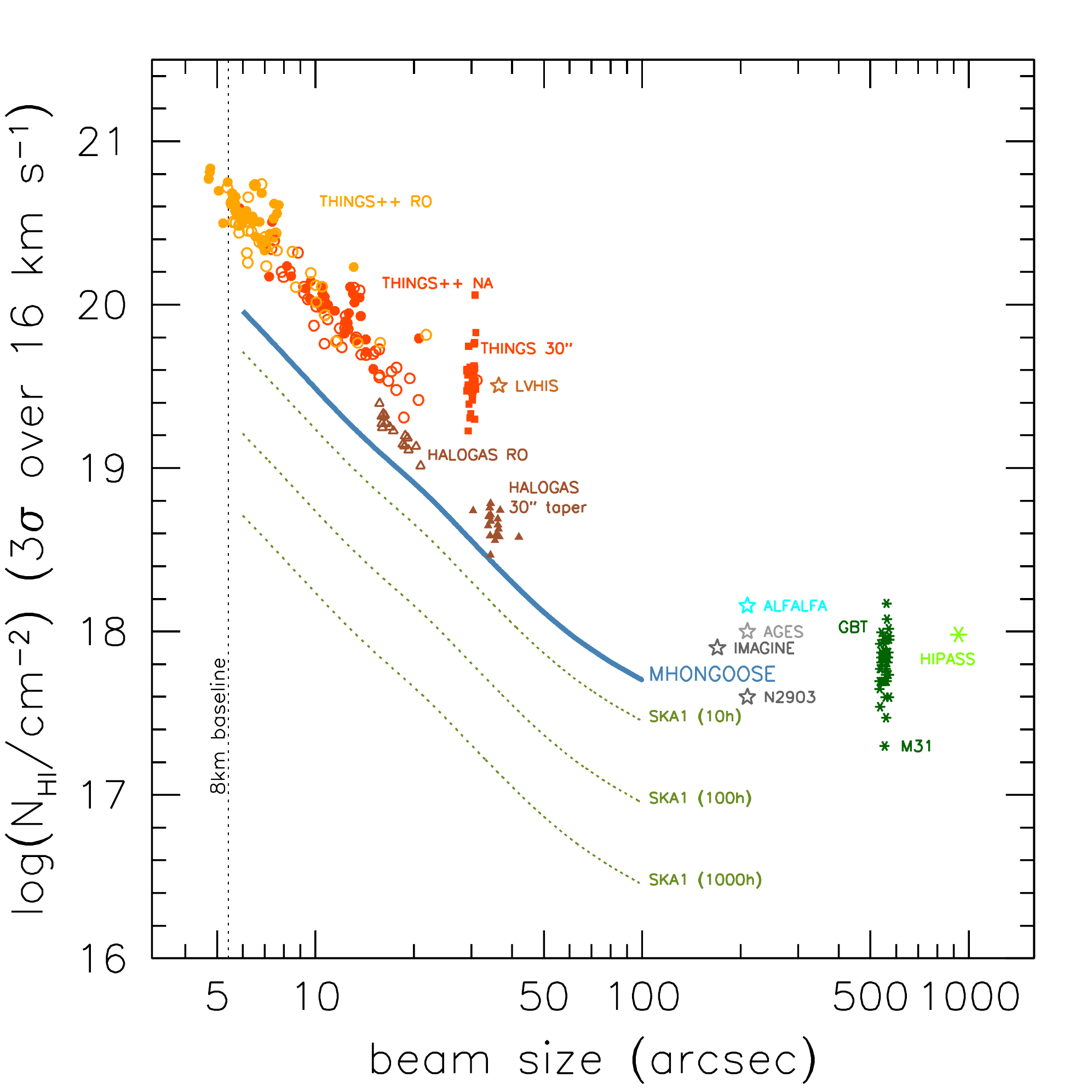}\\
\caption{MHONGOOSE \HI\ spectral line sensitivity versus other
  surveys. Light- and dark-orange filled circles show the
  sensitivities for individual galaxies in the THINGS survey, using
  {\tt robust=0.5} and natural weighting, respectively
  \cite{walter08}. Light- and dark-orange open circles show the same
  for the LITTLE THINGS survey \cite{hunter12}. Dark-orange squares
  show the natural-weighted THINGS observations spatially smoothed to
  $30''$. Open and filled brown trianges show the individual HALOGAS
  \cite{heald11} sensitivities at full resolution using {\tt
    robust=0}, and tapered with a $30''$ taper, respectively. Light-
  and dark-gray stars indicate observations of NGC 2903 by
  \cite{irwin09}, the AGES survey \cite{minchin10}, and the IMAGINE
  survey. Dark-green stars indicate the GBT observations of THINGS and
  HALOGAS galaxies by Pisano (in prep.). The bottom green star
  indicates the deep M31 observation by \cite{wolfe16} discussed in
  Sect.\ 4.2. Average sensitivities of the HIPASS, ALFALFA, and LVHIS
  surveys are also indicated. The MHONGOOSE sensitivity is indicated
  by the thick blue line.  The dotted green lines show the expected
  sensitivities for SKA1-MID for observing times of 10$^h$, 100$^h$
  and 1000$^h$.
  \label{fig:surveys}}
\vspace{-7pt}
\end{figure}

Most interferometric \HI\ surveys of nearby galaxies in the last decade
have concentrated on high angular resolution observations with a
fairly modest column density sensitivity. These are surveys such as
THINGS \cite{walter08}, LITTLE THINGS \cite{hunter12} and VLA-ANGST
\cite{ott12} which all reach column density limits of $\sim 10^{20}$
cm$^{-2}$.
The HALOGAS survey \cite{heald11} is an exception to this. It used
long integration times with the WSRT to reach column densities around
$\sim 10^{19}$ cm$^{-2}$, but at relatively low angular resolutions
($15''$--$30''$).  

To put MHONGOOSE in the context of these surveys, we compare their
respective sensitivities in Fig.\ \ref{fig:surveys}.  Sensitivities of
existing surveys have been taken from the source papers, and have all
been converted to a $3\sigma$ limit, integrated over a 16 km s$^{-1}$
channel. It is clear that the observations of THINGS and its siblings
only probe the high column density \HI. Smoothing these data to lower
resolutions increases their sensitivity somewhat, but the real jump in
sensitivity is made by the HALOGAS survey.  Recent, ultra-deep
observations of the THINGS and HALOGAS galaxies obtained with the GBT
\cite{pisano14, deblok14} are also indicated. These are some of the
deepest \HI\ observations ever done, but with a limited angular
resolution ($\sim 9'$). IMAGINE (PI A.\ Popping) is a survey currently
underway at the Australia Telescope Compact Array (ATCA) which also
aims to image low-column density structures around nearby galaxies
using the most compact configurations of ATCA. The angular resolution
will be a few times better than the GBT observations.

Also shown is the expected sensitivity of the MHONGOOSE
observations. It is abundantly clear that over the entire range of
angular resolution shown here the MeerKAT observations will be
superior\footnote{The MHONGOOSE column densities take into account the
  increased noise due to tapering to the desired resolution, as
  derived using the most recent MeerKAT antenna configurations.}.

At the highest resolutions, MHONGOOSE will achieve the resolution of
THINGS, but will be an order of magnitude deeper in column density. At
resolutions of $\sim 1'$ the observations reach the depth of the
deepest GBT observations of nearby galaxies, but with an order of
magnitude better angular resolution.

Using the known distances to the sample galaxies we can compare the
column density sensitivities as a function of physical scale.  For
each galaxy in the MHONGOOSE sample we can calculate the column
density as a function of physical resolution. As we vary the beam size
from $8''$ to $90''$, each galaxy creates a track in the column
density-physical resolution plane.

\begin{figure}[t]
\centering
\includegraphics[width=0.28\linewidth]{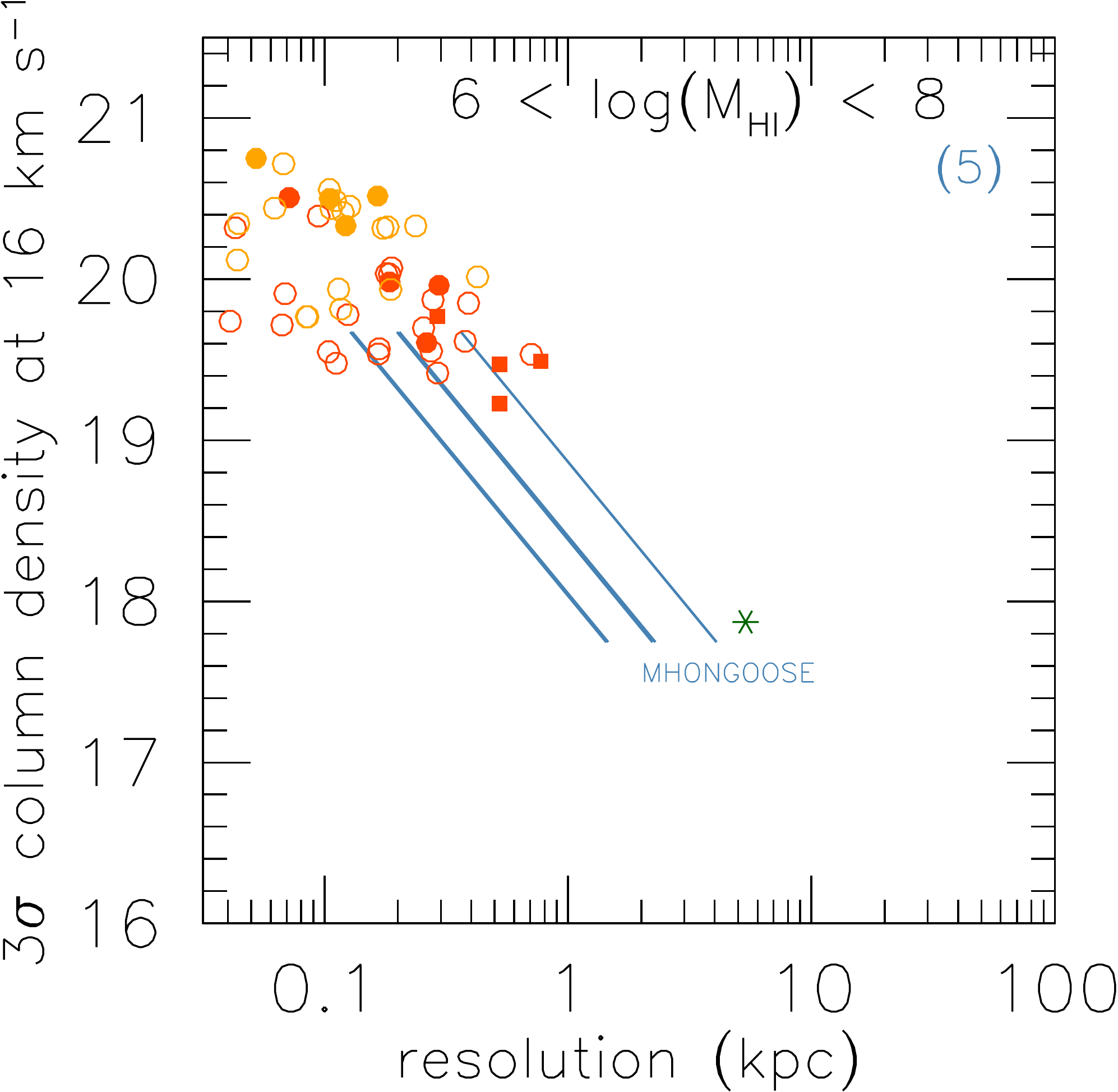}
\includegraphics[width=0.28\linewidth]{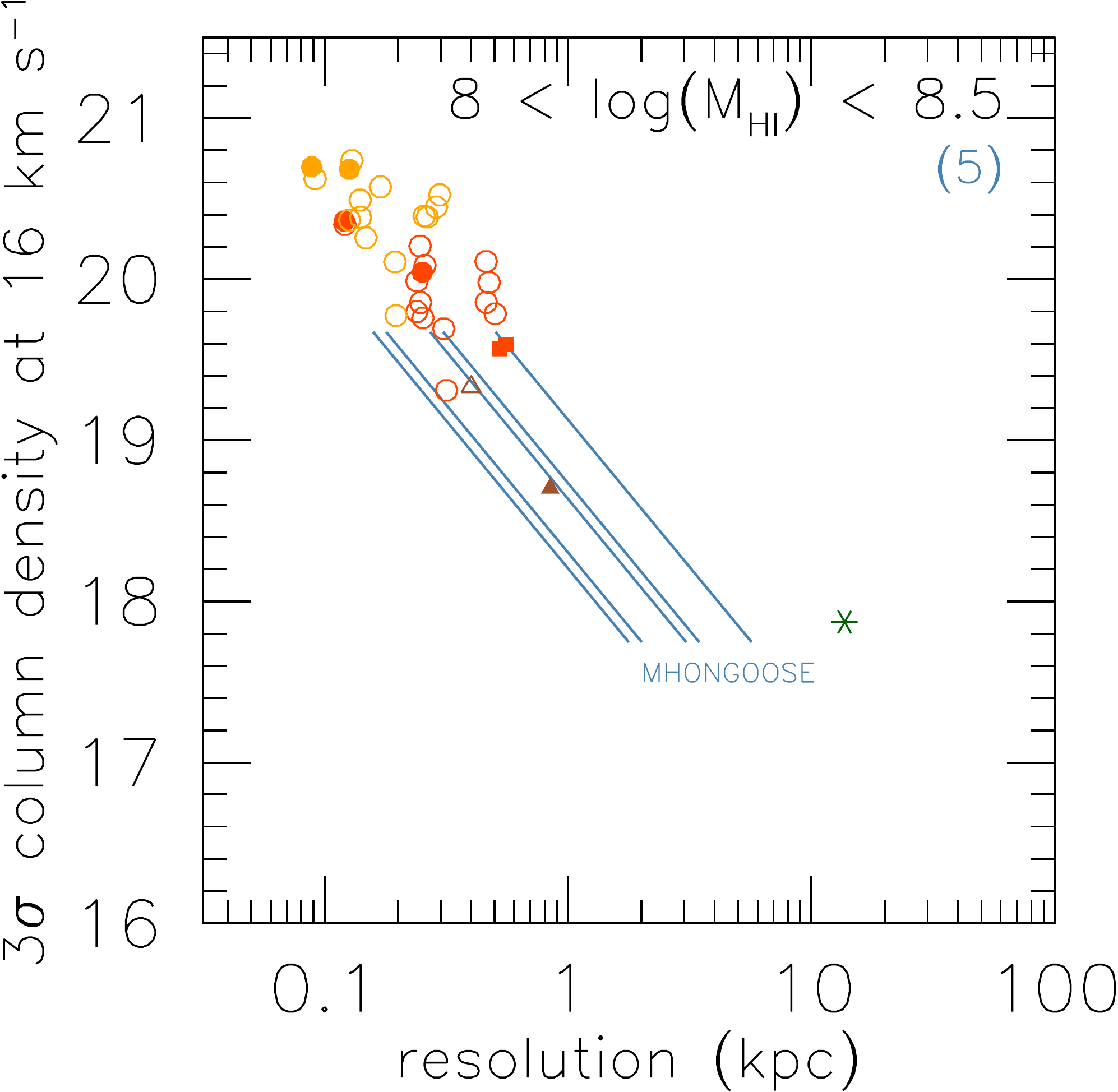}\includegraphics[width=0.28\linewidth]{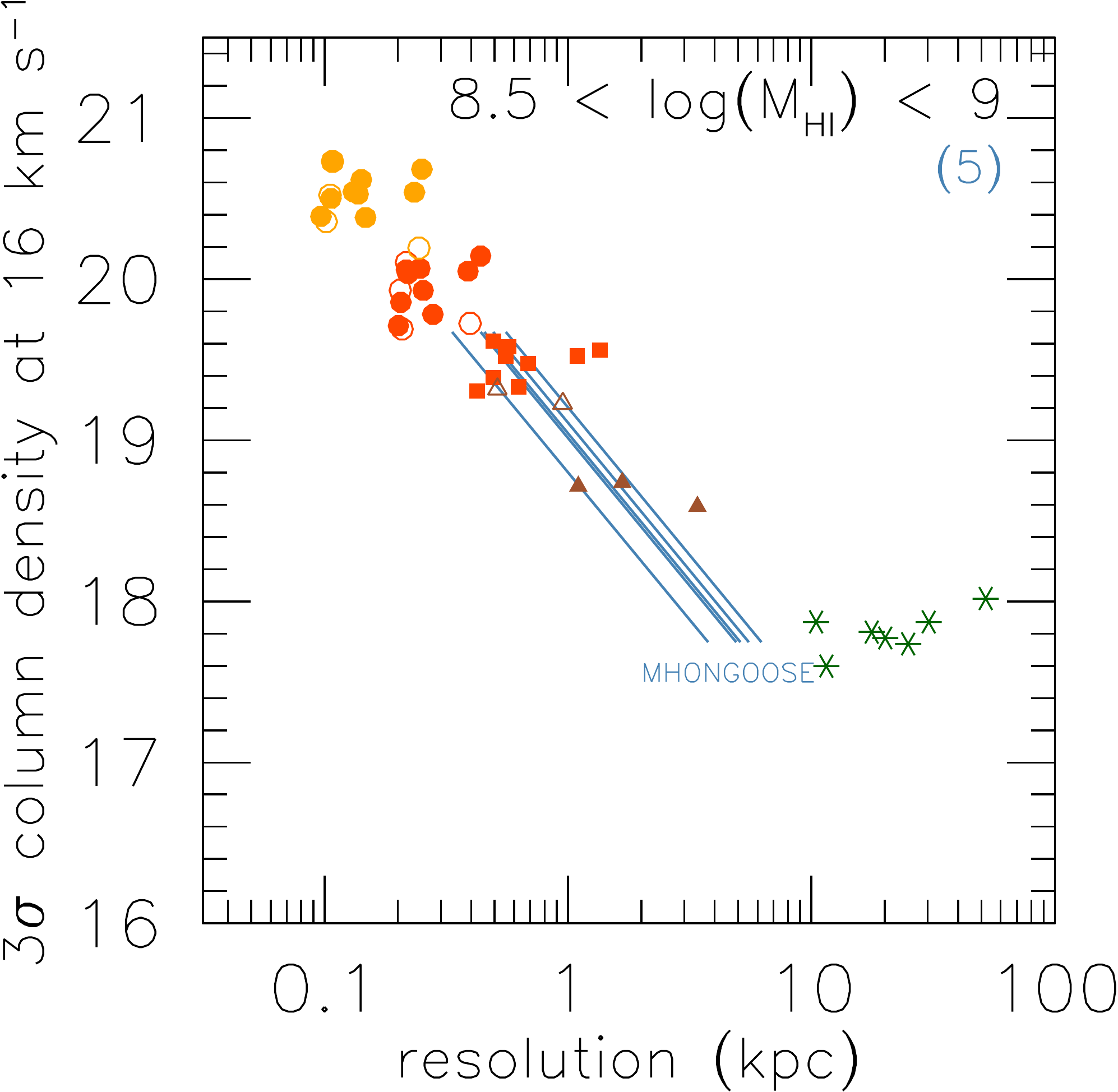}\\\vspace{5pt}
\includegraphics[width=0.28\linewidth]{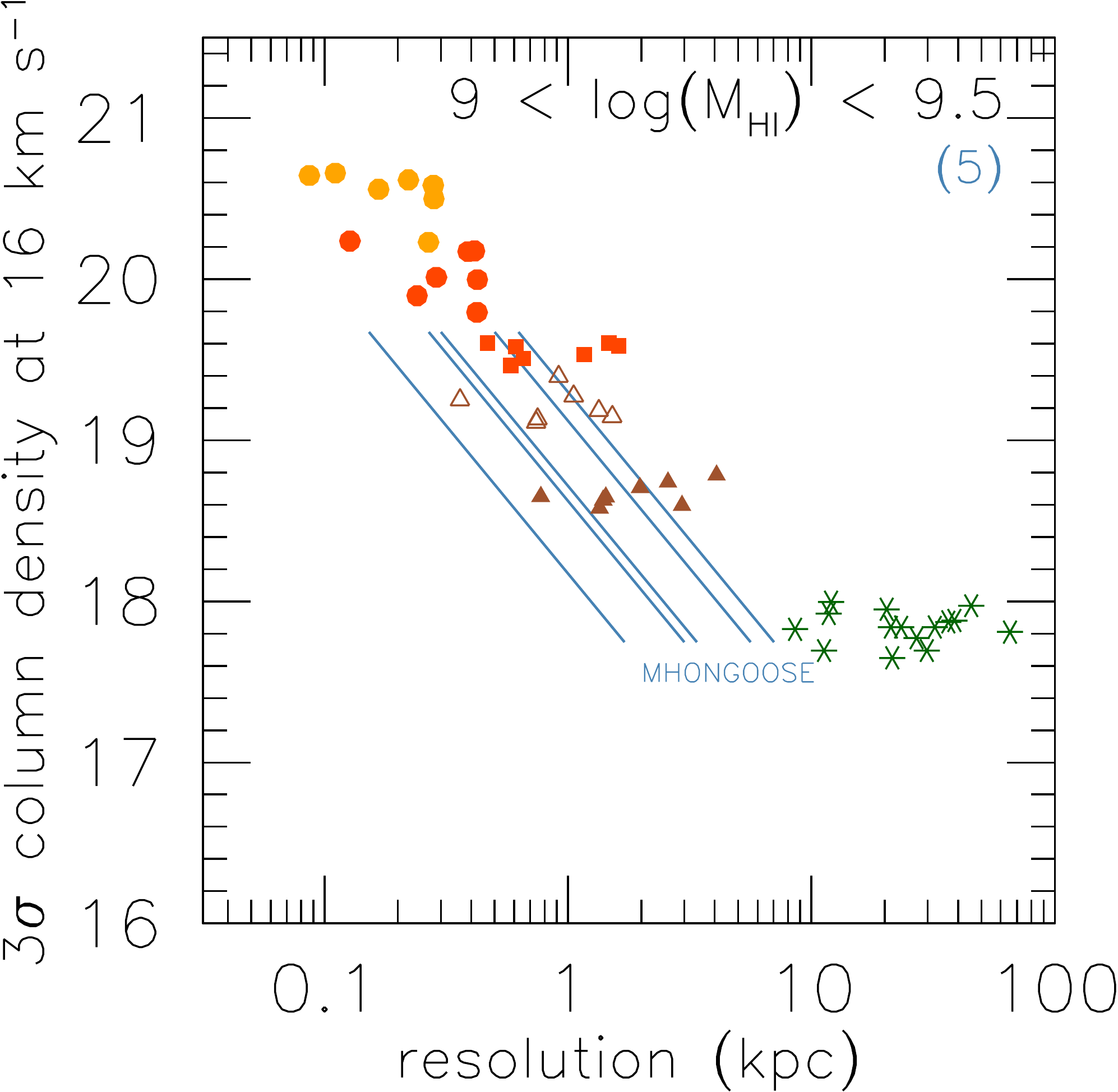}
\includegraphics[width=0.28\linewidth]{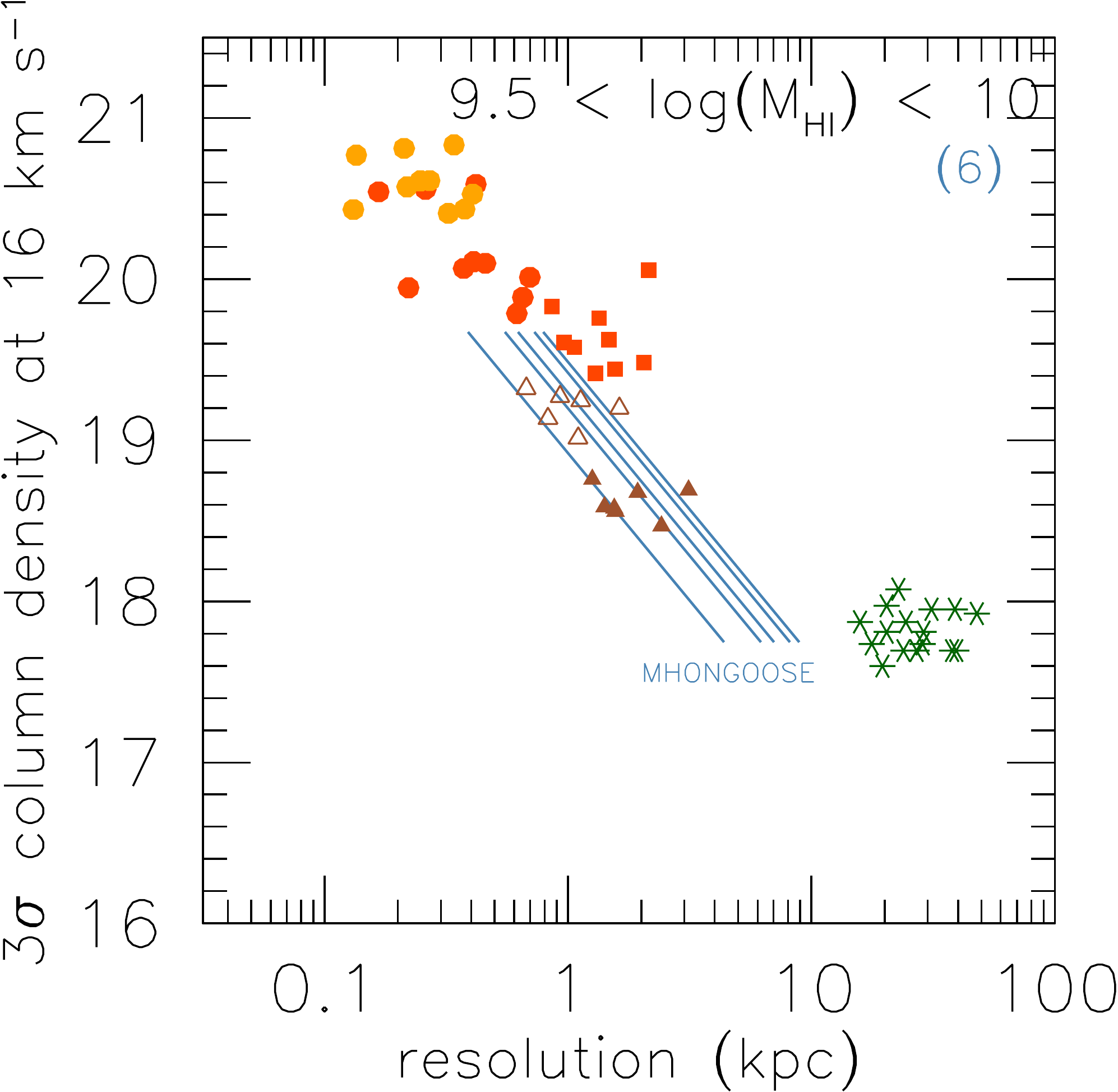}\includegraphics[width=0.28\linewidth]{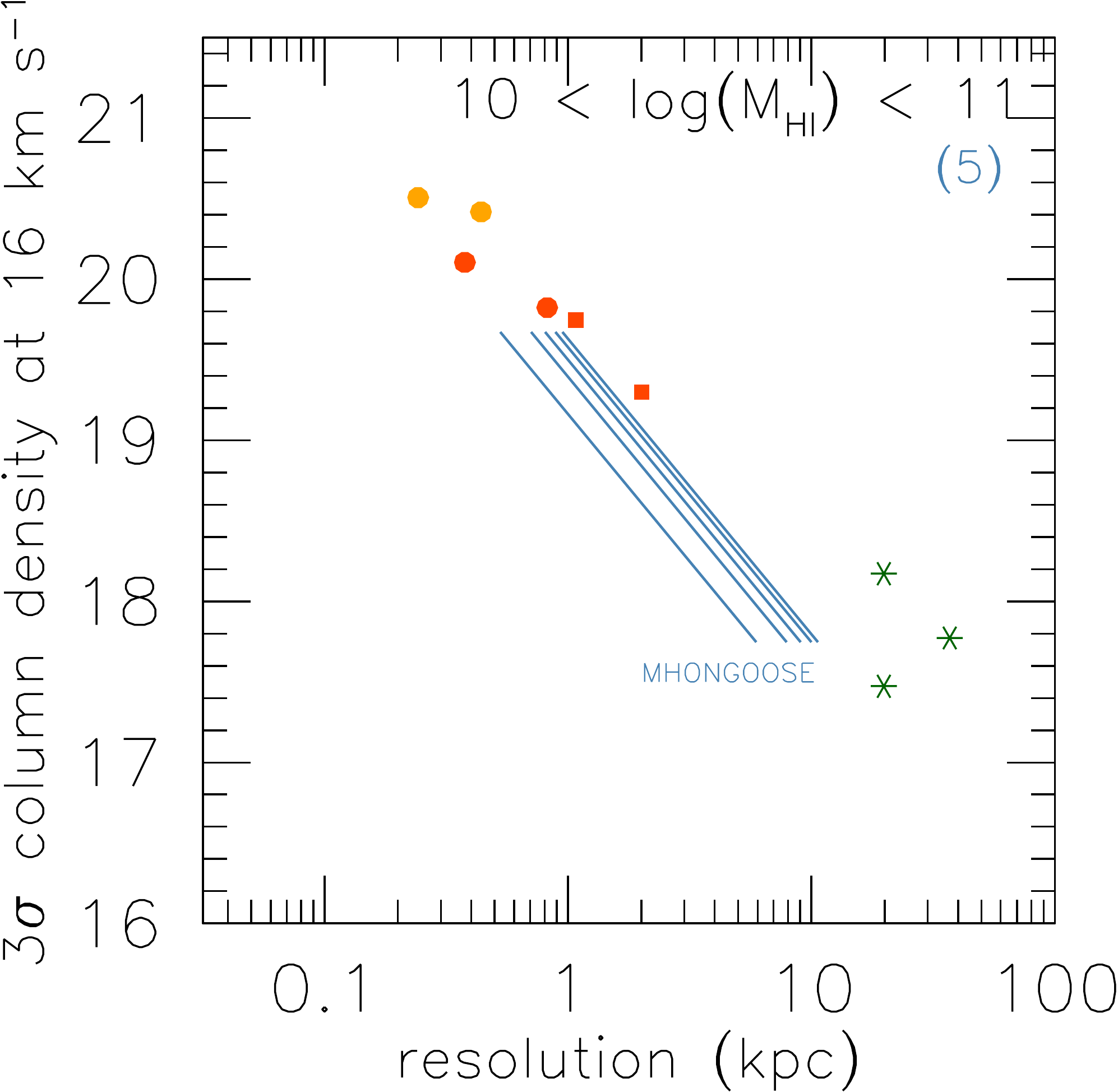}\\
\caption{Column density sensitivity for individual MHONGOOSE galaxies
  as a function of linear resolution. Symbols are as in
  Fig.\ \ref{fig:surveys}, and the blue lines show the sensitivity for each
  galaxy as the beam size is changed from $8''$ to $90''$. Panels show
  the galaxies in each \HI\ mass bin. Note that in the lower centre
  panel 6 galaxies are shown, as one of the MHONGOOSE pointings covers
  two galaxies at different distances.
\label{fig:surveykpc}}
\end{figure}

In Fig.\ \ref{fig:surveykpc} we show the tracks of the galaxies for
each of the \HI\ mass bins of the sample. We also show the galaxies from
the other \HI\ surveys with \HI\ masses in that same bin. We see that the
highest and lowest mass bins are very much unexplored at low column
densities. The galaxies of HALOGAS mostly fall within the
$9<\log(M_{HI})<10$ mass bin, but even here the MHONGOOSE observations
will probe an order of magnitude deeper for a given linear size.

\subsection{MHONGOOSE and low column densities}


Multiple independent lines of evidence show that the surface
area subtended by \HI\ at column densities near $10^{17}$ cm$^{-2}$ is a
factor of two larger than that seen at $10^{19}$ cm$^{-2}$
\cite{corbelli02, braun04, popping09, braun12}. In other words, sizes
of the \HI\ disks will increase in area by a factor of two compared to
the observations provided by surveys such as THINGS and HALOGAS.

\begin{figure}[t]
\begin{minipage}[t]{\linewidth}
\centering
\includegraphics[height=0.25\linewidth]{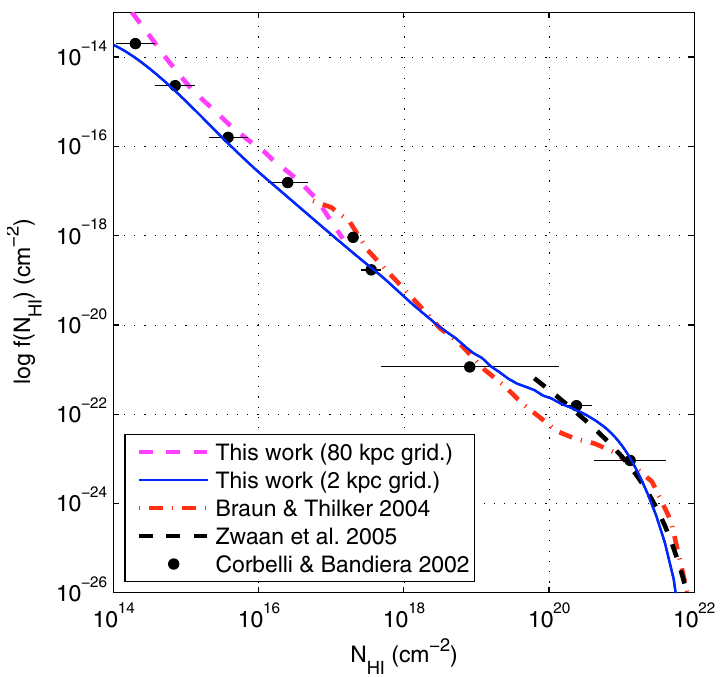}
\includegraphics[height=0.25\linewidth]{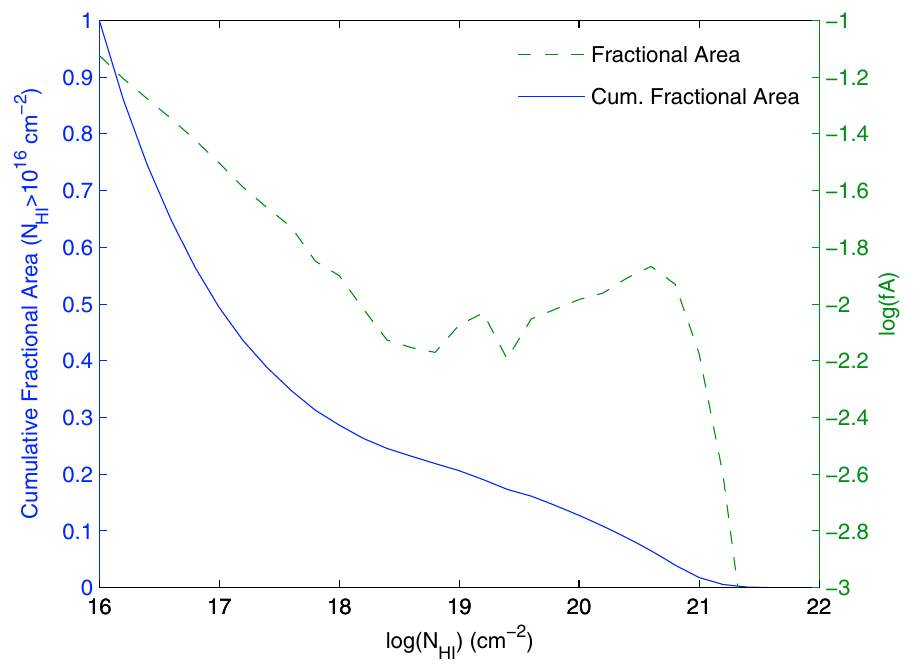}
\includegraphics[height=0.25\linewidth]{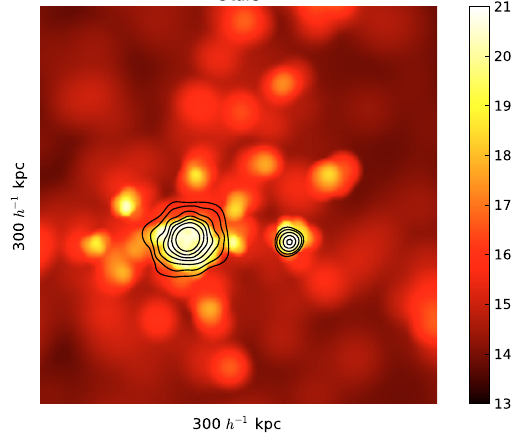}
\caption{Figures from \cite{popping09}. Left panel: the \HI\ column
  density distribution function. Note the dip near $10^{19}$ cm$^{-2}$
  causing an ``edge'' to the \HI\ disks of galaxies. Middle panel:
  fractional area (green) of a galaxy disk. The blue line shows the
  increase in area towards lower column densities. The area at
  $10^{16}$ cm$^{-2}$ has been arbitrarily set to 1. Right panel:
  simulation of the morphology of the low column density material. The
  material is expected to be clumpy. The bar indicates the column
  density as $\log(N_{HI}/\rm(cm^{-2}))$, contours indicate the locations of the stellar
  component.}
\label{fig:popping}
\end{minipage}
\vspace{-10pt}
\end{figure}

The left panel in Fig.\ \ref{fig:popping} shows the \HI\ column density
distribution function, or the likelihood that a line of sight
encounters a certain column density. This shows  that as one
goes to lower column densities, the area covered by the \HI\ increases,
{\it except} around $\sim 10^{19}$ cm$^{-2}$, where the slight
dip in the function leads to the observed ``edge'' of the \HI\ disk.

This is further illustrated in Fig.\ \ref{fig:popping} (center),
showing the fractional area $f_A = A(N_{HI})/{\rm d} \log(N_{HI})$ as
a function of column density $N_{HI}$. We see that the disk hardly
grows around $\sim 10^{19}$ cm$^{-2}$, but then increases in size
quickly again below a few times $10^{18}$ cm$^{-2}$. HALOGAS probed
the $10^{19}$ cm$^{-2}$ regime and, while it picked up low column
density gas in and around the disks of galaxies, it did not observe a
significant increase in the size of the disk.  MHONGOOSE will take us
in a regime where disk size growth is very pronounced.  The middle
panel in Fig.\ \ref{fig:popping} shows that the area subtended by the
$10^{17}$ cm$^{-2}$ emission is twice as large as that of the
$10^{19}$ cm$^{-2}$ emission and $\sim 65\%$ larger than that of the
$10^{18}$ cm$^{-2}$ emission.

However, detecting an increase in disk size is not the main goal of
these observations.  Rather, we want to characterize the morphology
and dynamics of the ultra-low column density material as this is not
expected to be in a smooth, homogeneously distributed
disk. Simulations by \cite{popping09} already indicate that the
$10^{17}$ cm$^{-2}$ material is likely distributed as clumps and
clouds of a few kpc in size (see the right panel in
Fig.\ \ref{fig:popping}). According to \cite{popping09},  these clouds are associated
with accretion from the cosmic web. In a recent study, \cite{wolfe16} have also shown
that the material detected between M31 and M33 (originally detected as
a smooth component by \cite{braun04}) is clumpy at similar scales
when observed at higher resolutions. This extremely deep observation is
indicated separately in Fig.\ 5.

Very deep single pointings with the GBT by D.J.~Pisano
(priv.\ comm.), arranged in a sparsely sampled grid around NGC 2403,
NGC 3198 and M31, and reaching column density limits of $\sim 10^{17}$
cm$^{-2}$, confirm this. Aside from the limited resolution, sparsely
sampled single pointings will, however, not be able to constrain the
dynamics and morphology of the lowest column density gas. MeerKAT is
currently the only telescope with the sensitivity and resolution that
can address this and therefore uniquely placed to investigate this
SKA1 key science question.


\subsection{Magnetic fields in the MHONGOOSE galaxies}

Magnetic fields are a crucial component of the ISM and the star
formation cycle, but their detailed properties and role in galaxy
evolution are still unclear. Observations of polarised synchrotron
radiation in nearby galaxies (see, e.g., \cite{beck15}) have given
us a clear picture of the overall structure and energetics of galactic
magnetic fields within the star forming disk. With modern radio
techniques, we are now opening the window to the detailed 3D structure
of gas and magnetic fields on scales relevant to constrain models of
ISM physics (e.g., \cite{mao15}). The relationship between
detailed magnetic structure and star formation will be probed for the
first time with the MHONGOOSE observations. We now appreciate that
there exist deep degeneracies between source models with different
combinations of synchrotron emission, magneto-ionic turbulence, and
Faraday rotation (e.g., \cite{horellou14}). Through broadband
synchrotron observations, these models can be distinguished
(e.g., \cite{heald15}, Williams et al., in prep.). The key is in
collecting polarimetric data over a wide range of $\lambda^2$ as even a
modest increase can lead to strong leverage on distinct models of
depolarisation effects. Techniques such as Rotation Measure Synthesis
and QU-Fitting will be employed to derive maximum benefit from the
MHONGOOSE polarimetry data.

For the nominal 0.9-1.6 GHz L-band continuum data, the instrumental
precision in Faraday Rotation Measure (RM) is 23 rad/m$^2$. This means
that for well-detected polarised emission (S/N$>8$) the effective RM
precision is $\lesssim1$ rad/m$^2$, sufficient to identify kpc-scale
fluctuations in magnetic field strength, within the star-forming ISM,
of order 20 nG (cf.\ the typical magnetic field strength of 1-10
$\mu$G). MHONGOOSE will provide detailed measurements of magnetic
field fluctuations in the ISM of the target galaxies.

MHONGOOSE will also provide a sensitive probe of the non-thermal
component in the outer parts of galaxies (beyond the main star forming
disk), tracing the large-scale morphology of magnetic fields and the
magnetic connection to the IGM. This will be possible both through
detection of diffuse polarised synchrotron radiation, as well as by
identifying foreground contributions to the RM of background polarised
radio galaxies. Thanks to the high sensitivity of MeerKAT's broadband
continuum mode (typical sensitivity of 0.75 $\mu$Jy beam$^{-1}$ for
48h on-source, or 0.15 $\mu$Jy beam$^{-1}$ for the ultra-deep
MHONGOOSE targets), diffuse synchrotron will be detected well outside the 25 mag arcsec$^{-2}$ diameter 
$D_{25}$, and a typical background polarised source density of
$\gtrsim 100$ deg$^{-2}$ will be recovered at S/N $\geq 10$ (based on
\cite{rudnick14}). This kind of density is sufficient to work out
the large-scale magnetic field properties even if no diffuse
synchrotron is detected from the foreground (target) galaxy itself \cite{stepanov08}.
The MHONGOOSE observations will thus provide new
constraints on the evolution of magnetic fields in galaxies, as well
as the possible magnetisation of the IGM \cite{beck15}. The MHONGOOSE
sample is an excellent testbed for addressing questions regarding the evolution of
magnetism, thanks to the broad diversity of galaxy properties that
results from the adopted sample selection (e.g., rotational velocity
and star formation rate, the two key ingredients of the galactic
dynamo process; see. e.g., \cite{beck96}).

\section{Summary}

MHONGOOSE is a MeerKAT Large Survey Project to map the neutral
hydrogen distribution in a sample of 30 nearby galaxies.  The sample
covers all inclinations, \HI\ masses from $\sim 10^6$ to $\sim
10^{11}\ M_{\odot}$, and luminosities from $M_R \sim −12$ to $M_R \sim
−22$. It samples the complete range of conditions found in local
galaxies: from prominent star forming disks all the way out to the
little-explored low-column density gas far out in the dark matter
halo. MHONGOOSE will be the first survey to provide a comprehensive
inventory of the processes driving the transformation and evolution of
galaxies in the nearby universe over 5 orders of magnitude in \HI\ mass
and column density.  The MHONGOOSE data, in combination with data at
other wavelengths, will provide the largest, most detailed and
versatile legacy database of nearby galaxy observations that will not
be surpassed until the SKA starts observing.

\end{document}